\documentclass[11pt,a4paper]{article}
\usepackage{jheppub}

\usepackage{dcolumn}
\usepackage{latexsym}

\newcommand{\be}{\begin{equation}}
\newcommand{\ee}{\end{equation}}
\newcommand{\bea}{\begin{eqnarray}}
\newcommand{\eea}{\end{eqnarray}}

\newcommand{\refc}[1]{(\ref{#1})}
\newcommand{\Ord}{\mathcal{O}}

\title{Probing boundary-corrections to Nambu-Goto open string energy levels in 3d $SU(2)$ gauge theory}

\author{Bastian B. Brandt}
\affiliation{Institut f\"ur Kernphysik, Johannes Gutenberg-Universit\"at Mainz,\\ Johann-Joachim-Becher-Weg 45, D-55128 Mainz}
\emailAdd{brandt@kph.uni-mainz.de}
\subheader{MKPH-T-10-26}

\abstract{
We measure the energy levels of the excitations of the flux tube
between static quark and antiquark in three-dimensional $SU(2)$ gauge theory.
Combining exponential error reduction techniques and a variational method we are able
to reduce the errors for the excited states significantly and to extract excited states in distinct
parity and charge conjugation channels.
It is conjectured that the infrared behavior (at large $q\bar{q}$ separation $R$)
of the flux tube is governed by an effective string theory.
Indeed previous simulations show good agreement between lattice data and predictions from
Nambu-Goto string theory.
Recently, new results on the effective string theory obtained corrections to the Nambu-Goto predictions
and showed that for the open string in three dimensions
first corrections should appears at order $1/R^4$.
They correspond to boundary terms in the worldsheet field theory.
These corrections are presumably small for the ground state, but significantly larger
for the excited states and lift the degeneracies of the free theory.
Assuming this functional form of the correction,
we obtain for the coefficient $b_2=-0.5(2)(2)$.
}

\arxivnumber{1010.3625}

\begin{document}

\maketitle

\section{Introduction}

The formation of a flux tube between static quark and antiquark in the QCD vacuum,
leads to a linearly rising potential and
is one of the possible mechanisms to describe quark confinement.
Simulations over the last years strongly support this picture (for a review see \cite{bali}).
It is conjectured that at large distances $R$ the dynamics of the flux tube
is governed by an effective string theory, describing the properties of strong interactions in the
infrared limit.
The construction of models for these QCD strings in terms of open bosonic string theories
have been attempted for a long time \cite{God}.
Today there are two different possible approaches for consistent effective string theories.
The first idea, due to Polchinski and Strominger \cite{PS}, was to construct an action consisting
of all terms invariant under conformal transformations that avoids the conformal anomaly in an arbitrary number of dimensions.
The first analysis of the closed string spectrum to $\Ord(R^{-1})$ was extended up to
and including $\Ord(R^{-5})$ in
\cite{drum1,drum2,pmd1,pmd2,pmd3,unpub,dass1,dass2,dass3}, showing that the spectrum is equivalent to the spectrum
of Nambu-Goto string theory (NG) \cite{Alvarez,Arvis}
\begin{equation}
 \label{eqarvis}
E^{NG}_{n}(R) = \sigma \: R \: \sqrt{ 1 + \frac{2\pi}{\sigma\:R^{2}} \: \left( n - \frac{1}
{24} \: ( d - 2 ) \right) } \; ,
\end{equation}
at least to that order.

The other approach is due to L\"uscher and Weisz \cite{LW-calc,LW-string}, using an action consisting of all possible terms
constructed directly from the transverse degrees of freedom. The corresponding coupling constants are fixed through
the duality between open and closed strings wrapping around a spatial dimension, which is valid
in the presence of at least one compactified spatial dimension of length $R$.
Recently their work was extended to even higher orders in $1/R$ and to the open string case \cite{ahar1,ahar2,r4cor}, using the $SO(1\mid d-1)$ Lorentz symmetry to constrain additional coupling constants (see also \cite{meyer}).
Leading order corrections to the closed string spectrum in $d>2+1$ possibly
appear at order $\Ord(R^{-5})$ for the excited states and at $\Ord(R^{-7})$ for the ground state.
For the energy spectrum of the open string the picture is different due to boundary terms in the effective action. Indeed it turns out that first corrections to all states may already appear at $\Ord(R^{-4})$,
also for $d=2+1$. In addition, there are possible corrections at $\Ord(R^{-5})$ for $d>2+1$.
It is thus natural to look for boundary corrections in the $d=2+1$ theory,
since these are the only ones contributing up to $\Ord(R^{-6})$ at least.

The predictions from these effective theories can be tested by comparison to simulations of pure Yang-Mills theories
and a lot of studies were done using different theories in $2+1$ and $3+1$ dimensions,
see \cite{Kuti} and references therein.
In this context it is important to note, that liftings of degeneracy in the open string energy levels
for $d=3+1$ in pure $SU(3)$ gauge theory where already reported in \cite{Juge1,Juge2,Juge3}.
For recent results on the closed string spectrum in $2+1$ and $3+1$ dimensional $SU(N)$ gauge theory see \cite{teper1,teper2}.
In the last years accurate measurements for the width of the flux tube became feasible as well
and results with improved systematics and high accuracy
coincide with the NG predictions for this quantity \cite{string-width1,string-width2}.

It is an ambitious project to compare the energy spectrum of the flux tube with the predictions
of the effective model.
The stringy behavior is expected to set in at relatively large
distances and the expected deviations from NG energy levels are small.
Thus one has to measure
a small effect at relatively large energies. In addition one has to control systematic effects as
e.g. contributions from excited states and effects coming from finite extent of the lattice.
Indeed most simulations so far haven't been able to reduce the errors and to control
the systematic effects sufficiently at the same time. This is especially true for open strings
with Dirichlet boundary conditions, corresponding to the excited states of a flux tube between
static quark and antiquark. The corresponding energy spectrum can be explored using Wilson loops,
but large temporal extent is needed to ensure a sufficient suppression of contaminations
from excited states.

In \cite{ownPos,BBPM} a new method was proposed to measure large loops with high accuracy,
based on the L\"uscher-Weisz multilevel algorithm \cite{LW-algo}, and has been applied
to the spectrum of the flux tube in three-dimensional $SU(2)$ gauge theory.
In this study this method is combined
with a variational method to extract excited state energies in a specific $(C,P)$-channel and reduce
excited state contaminations as much as possible.
This enables us to extract the energy levels accurate enough to compare
to the predictions from the effective theory.
Again we use $SU(2)$ gauge theory and work in three dimensions, but focus only on one
lattice spacing at $\beta=5.0$ with a relatively large lattice spacing in terms of $r_0$,
the Sommer scale \cite{Sommerpara}.
Simulating three instead of four dimensions is convenient because simulations are much faster
and $1/R^5$ corrections to the NG energy levels are absent in the three-dimensional theory.

The paper is organised as follows:
In the next section we begin by summarising the corrections to the Nambu-Goto energy levels
and lay the groundwork for the comparison to the measurements. In section \ref{ch-sim} we
turn to the details of the simulations, describe the extraction of the excited states
and the control of systematic errors. In section \ref{simresults} we discuss our results
and compare to the NG predictions. In section \ref{r4cor-ch} we compare to
the boundary corrections and finally draw our conclusions in the last section.

\section{Corrections to Nambu-Goto from boundary terms}
\label{corr}

In this section we discuss the boundary corrections
to the energy spectrum of open strings in $d=2+1$.
To keep things simple, we only discuss the main points and adopt the notation from \cite{ahar2,r4cor}.
As shown there, the only additional contribution to NG energy levels to $\Ord(R^{-5})$ in
the effective theory in $2+1$ dimensions is the contribution of a boundary term, leading to the Hamiltonian
\begin{equation}
\label{eqham}
H'_{2} = - b_2 \: \frac{\pi^{3}}{R^{4}} \: \left( 4 \sum_{n=1}^{\infty} n^{2} \: \alpha_{-n} \: \alpha_n + \frac{d-2}{60} \right) .
\end{equation}
The states of the Fock space contributing to the three lowest energy levels, denoted by $E_{n,i}$,
where $n$ is the energy level in the free theory and $i$ is the index for the degenerate states
at that level, are given by:
\begin{center}
\begin{tabular}{cl|r|c}
\hline
 energy & \hspace*{1mm} $\mid n,i \rangle \quad$ \hspace*{1mm} & \hspace*{1mm} representation \hspace*{1mm} & \hspace*{1mm} $(C,P)$ \hspace*{1mm} \\
\hline
\hline
$E_0$ & \hspace*{2mm} $\mid 0 \rangle$ & $ \mathbf{1} \: \mid 0 \rangle $ \hspace*{2mm} & $(+,+)$ \\
$E_1$ & \hspace*{2mm} $\mid 1 \rangle$ & $ \alpha_{-1} \: \mid 0 \rangle $ \hspace*{2mm} & $(+,-)$ \\
$E_{2,1}$ & \hspace*{2mm} $\mid 2,1 \rangle$ & $ \alpha_{-1} \: \alpha_{-1} \: \mid 0 \rangle $ \hspace*{2mm} & $(+,+)$ \\
$E_{2,2}$ & \hspace*{2mm} $\mid 2,2 \rangle$ & $ \alpha_{-2} \: \mid 0 \rangle $ \hspace*{2mm} & $(-,-)$ \\
\hline
\end{tabular}
\end{center}
Here $P$ is the parity quantum number of the state and $C$ what we call
the quantum number of charge conjugation
(called transverse and longitudinal parity respectively in \cite{teper2}).
To be precise: $C$ is the exchange between quark and antiquark, combined with a change in the direction
of the gluonic flow.
Using the commutation relations
\begin{equation}
\label{eqcomut}
\left[ \alpha_n , \: \alpha_{-n} \right] = n \: \delta_{nm}
\end{equation}
it is straightforward to compute the corrections to the NG energies $\epsilon_{n,i}$
of these states due to the Hamiltonian \refc{eqham}. The corrections are given by:
\begin{equation}
\label{eqcor}
\begin{array}{ccl}
\epsilon_0 & = & \displaystyle - b_2 \: \frac{\pi^{3}}{R^{4}} \: \frac{(d-2)}{60} \vspace*{2mm} \\
\epsilon_1 & = & \displaystyle - b_2 \: \frac{\pi^{3}}{R^{4}} \: \left( 4 + \frac{(d-2)}{60} \right) \vspace*{2mm} \\
\epsilon_{2,1} & = & \displaystyle - b_2 \: \frac{\pi^{3}}{R^{4}} \: \left( 8 + \frac{(d-2)}{60} \right) \vspace*{2mm} \\
\epsilon_{2,2} & = & \displaystyle - b_2 \: \frac{\pi^{3}}{R^{4}} \: \left( 32 + \frac{(d-2)}{60} \right)
\end{array}
\end{equation}

As we see, already the ground state obtains a $\Ord(R^{-4})$ correction from the boundary terms,
which is nevertheless strongly suppressed by a factor $1/60$ and therefore hardly visible in
our simulation, except for the unlikely case that $b_2$ is large.
Nevertheless it might be visible in high accuracy simulations for the ground state,
as e.g. performed in \cite{PundD1,PundD2}.
The corrections to the excited states are enhanced by a factor 240 or more compared to the ground state
correction and are therefore much easier to detect,
even though it is harder to extract these states numerically.
In addition the degeneracy of the states at $n=2$ in the free theory is lifted
and the magnitude of the corrections to the two formerly degenerate states
differ by roughly a factor of 4.

Important for the comparison between measurements and the predictions is the radius of convergence
of the expansion of the string energy levels in $1/R$,
which can be estimated by $x_c=1/\lambda$,
the radius of convergence for the expansion of
$\sqrt{1+\lambda\:x}$ around $x=0$.
In the effective string theory the radius of convergence corresponds to a critical length
$R_c\:\sqrt{\sigma}=\sqrt{1/x_c}$ below which the expansion of the square root ceases to be convergent.
For closed strings this critical length is relatively large, $R_c\:\sqrt{\sigma}\gtrsim 3.4$ for $d=2+1$,
while for the open string we have
\begin{equation}
\label{eqradconv}
\left. R_c\:\sqrt{\sigma} \right|_{n=1} = 2.45 \: , \qquad \left. R_c\:\sqrt{\sigma} \right|_{n=2} = 3.51 \; .
\end{equation}
As can be seen from \cite{BBPM,Pushan2a,Pushan2b}, extracting the energies of a flux tube of this length
is possible, especially for the first excited state.
Nevertheless the coefficient $b_2$ is not known {\it ab initio} and we have to extract it from the
data. We discuss the extraction of $b_2$ and the results in section \ref{r4cor-ch}.

\section{Details of the simulations}
\label{ch-sim}

\begin{figure*}[t]
\centering
\begin{minipage}[c]{0.45\textwidth}
\centering
\includegraphics[width=\textwidth]{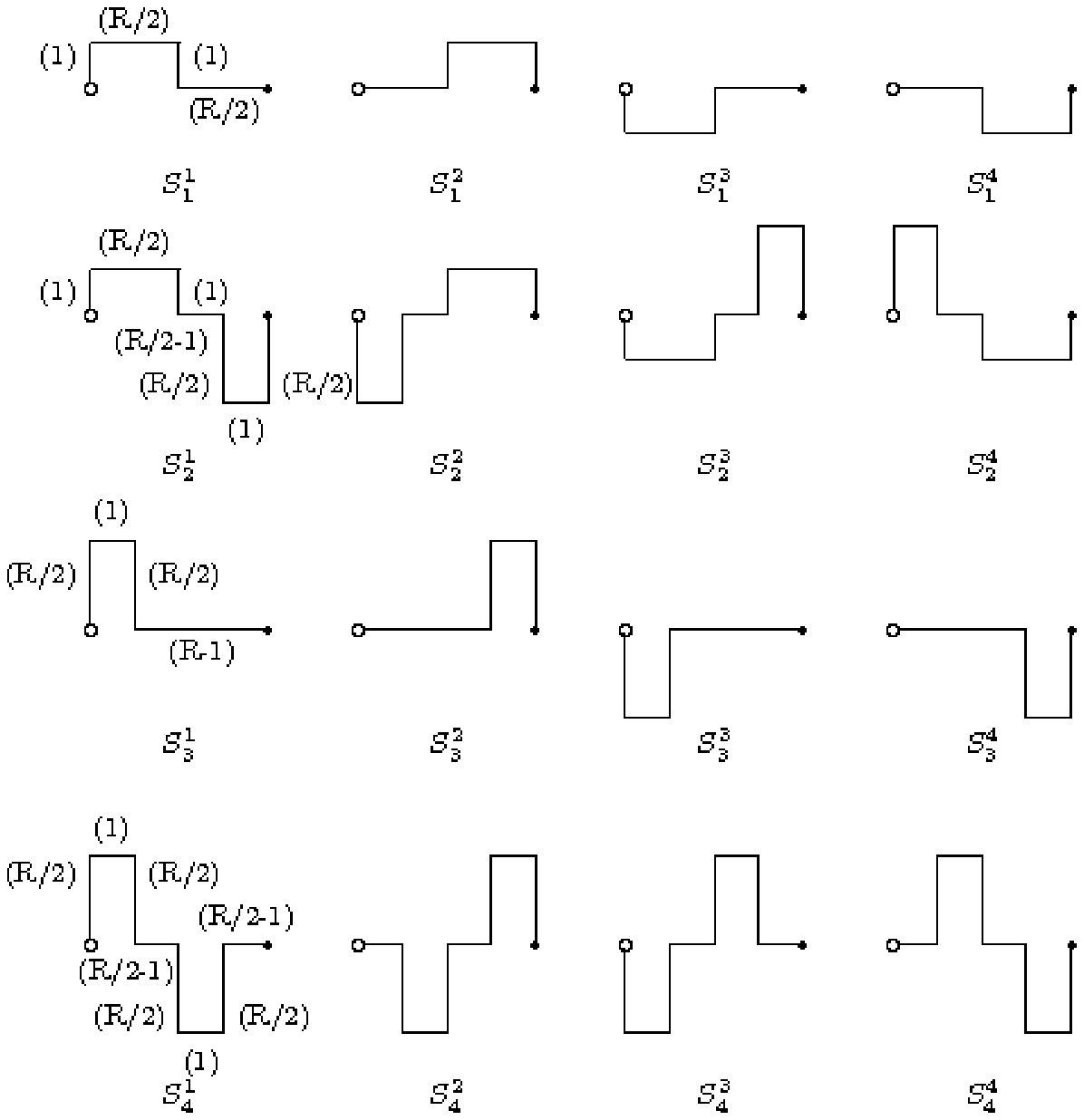}
\end{minipage}
\begin{minipage}[c]{0.45\textwidth}
\centering
\includegraphics[width=\textwidth]{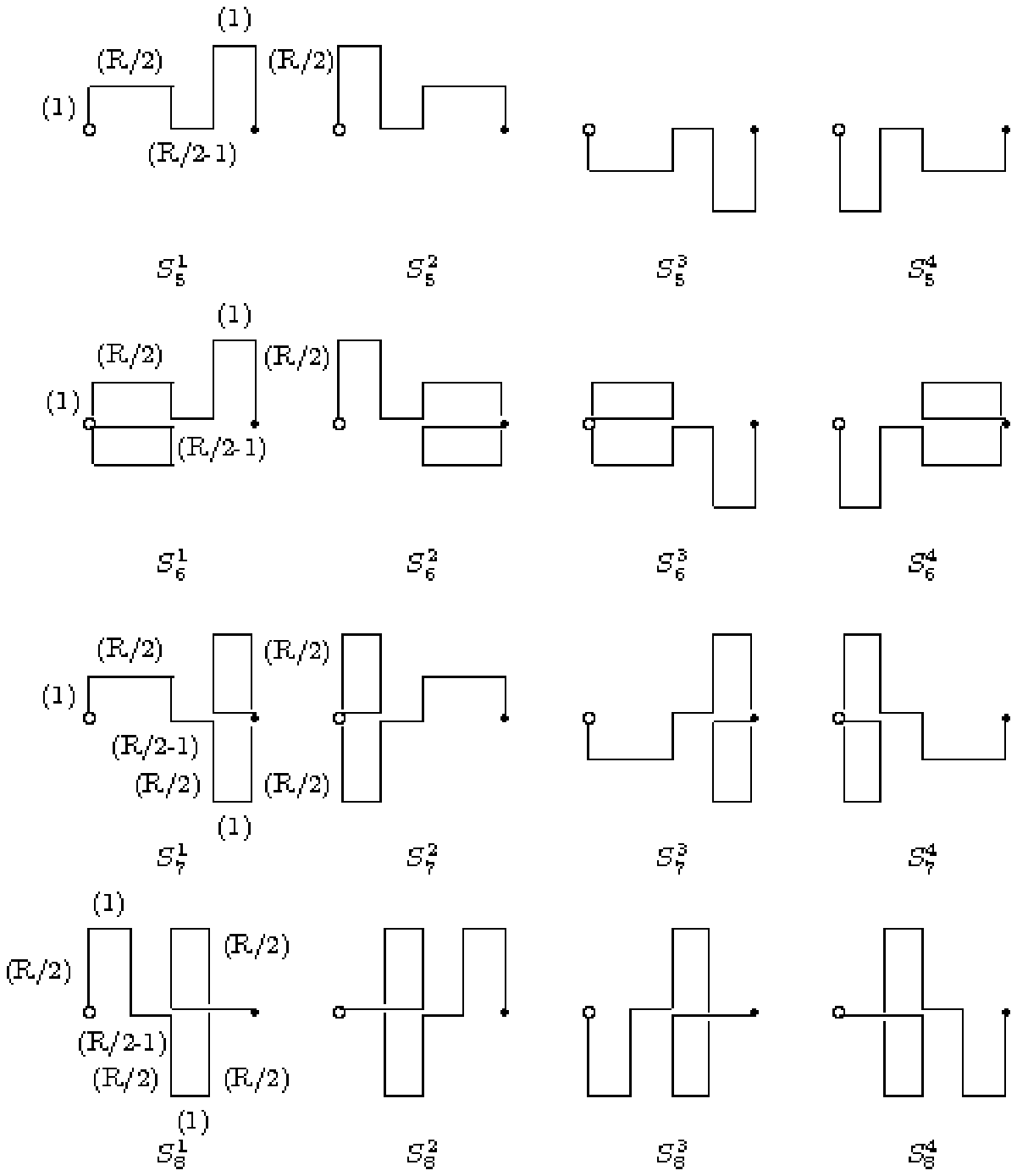}
\end{minipage}
\caption{Sets of operators to construct the correlation matrices in different $(C,P)$ channels.}
\label{fig1}
\end{figure*}

Wilson loops are the observables suited to consider the
excited states of the flux tube and were already used in \cite{Juge1,Juge2,Juge3} and
\cite{ownPos,BBPM,Pushan2a,Pushan2b} to extract
the spectrum of the excitations.
In this study, we combine the use of correlation matrices with the error
reduction method from \cite{ownPos,BBPM}
to extract the excitations spectrum in the $(C,P)$-channels from large Wilson loops
and suppress contributions from other
excited states as much as possible.

\subsection{Extraction of the energy spectrum}
\label{extraction}

\subsubsection{Extraction of eigenvalues}

To extract the energy spectrum we use 8 different operator sets, shown in figure \ref{fig1}.
From these sets we first construct $(C,P)$-projectors,
given by the linear combinations (see also \cite{BBPM})
\begin{equation}
\label{eigstates}
\begin{array}{l}
S_{i}^{++} = S_{i}^{1} + S_{i}^{2} + S_{i}^{3} + S_{i}^{4} \\
S_{i}^{+-} = S_{i}^{1} + S_{i}^{2} - S_{i}^{3} - S_{i}^{4} \\
S_{i}^{--} = S_{i}^{1} - S_{i}^{2} - S_{i}^{3} + S_{i}^{4} \\
S_{i}^{-+} = S_{i}^{1} - S_{i}^{2} + S_{i}^{3} - S_{i}^{4} \; .
\end{array}
\end{equation}
With these projectors we construct a correlation matrix in each of the $(C,P)$-channels.

To obtain the energies one has to extract the eigenvalues in the limit $T\to\infty$.
Extracting these eigenvalues is a delicate topic, since loops with a temporal extent much larger than
$3\:r_0$ are out of reach with todays algorithms.
In addition the number of operators $N$ used to construct correlation matrices
is finite, leading to the problem that even though one has extracted
the eigenvalues carefully there might be a remaining mixing with states belonging to other energy levels.
Usually one would use a generalised eigenvalue problem (GEVP), as discussed e.g. in \cite{forcrand}.
The GEVP has the advantage that the excited states contributing to each of the eigenvalues start at $n=N+1$
as shown in \cite{GEVP}.
In our case it is hardly possible to establish this method accurately,
since the correlation matrix might be ill-conditioned even for the smallest values of $T$.
This is especially an issue in the case of the correlation matrices in the $(-,-)$ and $(-,+)$-channels,
where the signal to noise ratio of the larger eigenvalues becomes very small.
We therefore diagonalise the correlation matrices at each value of $T$ separately to obtain the
eigenvalues $\lambda^{CP}_{n}(R,T)$ with $n=0,\ldots,7$, using the $QR$ reduction method (see e.g. \cite{golub})
and check explicitly the $T$ dependence of the overlaps between eigenvalues and operators.
This method is analogous to the method used in \cite{teper2}, where a variational criterion has been
applied at each value of $T$ seperately.
For all states discussed in this study there is no significant change in the overlaps in the range of
$4\leq T \leq 12$ and we do not expect to see any relicts of this in the data.
In contrast to the case of the GEVP we now have a remaining mixing with each energy level in the channel,
even though the mixing between the smallest eigenvalues should be suppressed if we have chosen
the operators wisely.
In \cite{teper1} and \cite{teper2} this mixing is minimised by using a very large number of basis states.
Increasing the number of operators above a limit of around 50 is
hardly possible in combination with our error reduction algorithm,
since the needed computing ressources grows drastically.

\subsubsection{Groundstate energies in a given channel}

In this study we mainly focus on the ground states of the $(C,P)$-channels, except for
the first excited state in the $(+,+)$-channel, which is discussed in the next section.
Each of the eigenvalues corresponding to the ground state in a channel now obeys the spectral representation
\begin{equation}
\label{eigst2}
 \lambda^{CP}_{0}(R,T) = \sum_{j=0}^{\infty} \beta^{CP}_{j}(R) \: e^{-E^{CP}_{j}(R)\:T} \; ,
\end{equation}
where $E^{CP}_{j}$ are the energies in the $(C,P)$-channel
and $\beta^{CP}_{j}$ the overlaps with the eigenstate.

If the excited states are suppressed sufficiently, we can extract the
corresponding energies using the asymptotic behavior
\begin{equation}
\label{eqasympt}
-\ln\left(\lambda^{CP}_{0}(R,T)\right) = \bar{E}^{CP}_{0}(R)\:T - \ln\left(\beta^{CP}_{0}(R)\right) \; ,
\end{equation}
which is exact in the limit $T\to\infty$.
In practice $\bar{E}^{CP}_{0}(R)$ is obtained by fitting the data
to \refc{eqasympt} together with the logarithm as an additional fit parameter.
These energies are called na\"ive from now on, because contaminations from excited states are neglected.

Unfortunately the simulations have shown that the suppression of the excited states is not sufficient
with the loop sizes in reach with the present methods of error reduction.
This is especially crucial for larger values of $R$ as shown in section \ref{simresults}.
It is therefore mandatory to take the corrections into account. This can be done by fitting the
eigenvalues for different temporal extents $T_a$ and $T_b$ to the leading order formula (see also \cite{Pushan2a,Pushan2b,ownPos,BBPM})
\begin{equation}
\label{eqfit1}
\begin{array}{rl}
\displaystyle  - \frac{1}{T_{b}-T_{a}} \: \ln \left[ \frac { \lambda^{CP}_{0} ( R , T_{b} ) } { \lambda^{CP}_{0} ( R , T_{a} ) } \right] = & \displaystyle E^{CP}_{0} ( R ) + \frac{1}{T_{b}-T_{a}} \: \alpha^{CP}(R) \: e^{ - \delta^{CP}(R) \: T_{a} } \vspace*{2mm} \\
 & \displaystyle \times \left( 1 - e^{ - \delta^{CP}(R) \: ( T_{b} - T_{a} ) } \right) \: .
\end{array}
\end{equation}
Here $T_a<T_b$, $\displaystyle \alpha^{CP}(R)=\beta^{CP}_1(R)/\beta^{CP}_0(R)$ and $\delta^{CP}(R)$ is the energy gap
to the first excited state in the channel.
The energies $E^{CP}_{0} ( R )$ are called
corrected from now on and are obtained as fit parameters together with the $\alpha$'s and $\delta$'s
from fits to all possible combinations of $T_a$ and $T_b$. The control of these fits is discussed in more detail
in section \ref{chfit}.

\subsubsection{Excited states in a given channel}

For an eigenvalue corresponding to an excited state in a channel the situation is different due to possible
mixings with states of smaller energy. The only excited state in a channel used in this study is the
first excited state in the $(+,+)$-channel on which we focus in this section.
The only state with smaller energy contributing is the corresponding ground state,
whose energy $E^{++}_0(R)$ can be extracted with the procedure described above.
For the na\"ive energy $\bar{E}^{++}_1$ we can use a fit to the form \refc{eqasympt}, since mixings with other energy
levels are neglected in this quantity. Nevertheless the extraction of the corrected energy $E^{++}_1$
needs some additional discussion.

A fit to the form \refc{eqfit1} removes the contaminations from states with larger energy values.
This remains true for excited states in a channel, but one has to include also the corrections coming from
states with smaller energy. For the state $E^{++}_1$ we can check the contribution of the state $E^{++}_0$
by adding the corresponding term
\begin{equation}
\label{corr-lightstate}
\frac{\gamma(R)}{T_{b}-T_{a}} \: e^{ (E^{++}_1(R)-E^{++}_0(R)) \: T_{a} } \: \left( 1 - e^{ (E^{++}_1(R)-E^{++}_0(R)) \: ( T_{b} - T_{a} ) } \right)
\end{equation}
to the fit \refc{eqfit1}.
Performing a fit using \refc{eqfit1} together with \refc{corr-lightstate} shows, that the contribution
of $E^{++}_0(R)$ should be well below our statistical errors, since $\gamma$ is in all cases
a number below $10^{-4}$, so that the total contribution of \refc{corr-lightstate}
is below $5\cdot10^{-3}$, well below the statistical errors for the corrected
energy $E^{++}_1$.
The other fit parameters agree well with the parameters obtained by a fit to the form
\refc{eqfit1} within the statistical errors.
As corrected values for $E^{++}_1(R)$ we thus use the results from a fit to the form \refc{eqfit1}
and neglect the contribution of the ground state.

\subsubsection{Energy differences}

The extraction of the energy differences can be treated independently from
the extraction of the total energies. In this way the energy differences might serve as an independent
check of the asymptotic results for the total energies.
Using \refc{eigst2} one can obtain for the na\"ive energy differences the formula
\begin{equation}
\label{eqasympt2}
-\ln\left(\frac{\lambda^{CP}_{n}(R,T)}{\lambda^{CP'}_{m}(R,T)}\right) = \left[\bar{E}^{CP}_{n}(R)-\bar{E}^{CP'}_{m}(R)\right]\:T - \ln\left(\frac{\beta^{CP}_{n,0}}{\beta^{CP'}_{m,0}}\right) \; ,
\end{equation}
which is similar to \refc{eqasympt} for the total energies.
This formula for na\"ive differences is also valid for the differences between excited states
in a given $(C,P)$-channel.

As before we are mainly interested in the differences between the energies of the ground states
in different channels. For these the leading order formula for the corresponding corrected
energy difference is the same as in \cite{BBPM},
\begin{equation}
\label{eqfit2}
\begin{array}{rl}
- & \displaystyle \frac{1}{T_{b}-T_{a}} \: \ln \left[ \frac { \lambda^{CP}_{0} ( R , T_{b} ) \: \lambda^{CP'}_{0} ( R , T_{a} ) } { \lambda^{CP}_{0} ( R , T_{a} ) \: \lambda^{CP'}_{0} ( R , T_{b} ) } \right] \vspace*{2mm} \\ = & \displaystyle \left[E^{CP}_0(R)-E^{CP'}_0(R)\right]
+ \frac{1}{T_{b}-T_{a}} \: \bar{\alpha}(R) \: e^{ - \bar{\delta}(R) \: T_{a} } \left( 1 - e^{ - \bar{\delta}(R) \: ( T_{b} - T_{a} ) } \right) \; ,
\end{array}
\end{equation}
where $\bar{\alpha}$ is a suitable combination of the overlaps and $\bar{\delta}$ corresponds
to the energy gap to the next excited state in the channel $CP'$
(which is equivalent to the gap in the channel $CP$ to leading order in $1/R$).
In addition to the energy differences between the ground state energies we are also interested
in the energy differences between $E^{++}_1$ and $E^{++}_0$. In this case
the two terms corresponding to the mixing between $E^{++}_1$ and $E^{++}_0$ are a
term of the form \refc{corr-lightstate} and a term of the form
\begin{equation}
\label{eqdiffitcor}
\begin{array}{l}
\displaystyle \frac{\bar{\alpha}(R)}{T_b-T_a} \: \left(\frac{\gamma(R)}{\beta_0(R)}\right) \: e^{(E^{++}_1(R)-E^{++}_0(R)-\bar{\delta}(R))\:T_a} \vspace*{2mm} \\
\displaystyle \times \left(1-e^{(E^{++}_1(R)-E^{++}_0(R)-\bar{\delta}(R))\:(T_b-T_a)}\right) \ll 10^{-4}
\end{array}
\end{equation}
The total contribution of these terms is below $5\cdot10^{-3}$ and thus negligible
compared to the statistical errors.
We therefore use a fit to the form \refc{eqfit2} for that difference, too.

\subsection{Algorithm and simulation parameters}

The need for suppression of the contributions from excited states and the onset of
string-like behavior in the large $R$ regime
demands large Wilson loops, and an efficient algorithm with sufficient error reduction for large loops
is needed to extract the spectrum with high accuracy. The algorithm used in this study is a variation
of the L\"uscher Weisz algorithm \cite{LW-algo} and discussed in detail in \cite{BBPM}.

Our simulations were done using the $SU(2)$ Wilson plaquette action in $2+1$ dimensions at $\beta=5.0$.
The Sommer parameter is known e.g. from \cite{ownPos,Pushan2a,Pushan2b}
and listed together with the simulation parameters in table \ref{tab1}.
In fact the lattice spacing in terms of the Sommer parameter is not very small,
but a comparison from the results of the four
different $\beta$ values from \cite{ownPos} and \cite{BBPM} shows that there is not much movement
in the energies expressed in terms of $r_0$ over the whole range of $5.0\leq\beta\leq12.5$.
We therefore expect that our results are relevant for the continuum as well.
In contrast to \cite{Pushan2a,Pushan2b,ownPos,BBPM}, we work with 5 different temporal extents instead of 4,
which leads to strong improvements in the extraction of the asymptotic behavior and the control of
contaminations from excited states via \refc{eqfit1} and \refc{eqfit2}.

\begin{table*}[t]
\centering
\begin{tabular}{ccc|ccccccc}
\hline
$\beta$ & \hspace*{2mm} $r_0/a$ \hspace*{2mm} & \hspace*{1mm} $R$ \hspace*{1mm} & \hspace*{1mm} $T$ \hspace*{1mm} & \hspace*{1mm} $T/r_0$ \hspace*{1mm} & \hspace*{1mm} $t_s$ \hspace*{1mm} & lat size & \hspace*{1mm} $N_s$ \hspace*{1mm} & \hspace*{1mm} $N_t$ \hspace*{1mm} & \# meas \\
\hline
\hline
5.0 & 3.9536(3) & 4-12 & 4 & 1.01 & 2 & $32^3$ & 16000 & 1500 & 3200 \\
 & & & 6 & 1.52 & & $36^3$ & & 2000 & 3200 \\
 & & & 8 & 2.02 & & $40^3$ & & 6000 & 5100 \\
 & & & 10 & 2.53 & & $40^3$ & & 12000 & 6400 \\
 & & & 12 & 3.04 & & $48^3$ & & 16000 & 8600 \\
\hline
\end{tabular}
\caption{Run parameters of the simulations. $r_0$ is the Sommer parameter, $t_s$ the temporal extent of the sublattices in the LW algorithm, $N_s$ the number of updates of the sublattice containing the spatial operators and $N_t$ the number of updates of the sublattices containing only the time transporters.}
\label{tab1}
\end{table*}

Statistical errors are estimated using the usual binned jackknife method
with 50 bins for all measurements.
We explicitly checked that none of the error estimates varies more than a few percent with bin size.

\subsection{Sources of systematic effects}

\subsubsection{Finite volume effects}

The first class of sources of systematic errors in addition to the ones discussed in section
\ref{extraction} is due to finite extent $L$ of the lattice.
Here periodic boundary conditions are applied to the lattice, making it possible for the
loops to interact with themselves by around-the-world glueball exchanges.
In the string picture these exchanges corresponds to handles on the world sheet,
wrapping around a compactified dimension.
The contribution of such a handle to the signal of the Wilson loop takes the form
\begin{equation}
\label{eqfinitesize}
a(L')\:\exp\left(-m_G\:L'\right) \; ,
\end{equation}
where $L'$ is the length of the handle, $m_G$ the mass of the lightest glueball
(measured in \cite{miketeper} for $SU(2)$ in $d=2+1$ and at $\beta=5.0$)
and $a(L')$ is the overlap with the loop.
In principle these handles can wrap around any of the directions, but it is immediately clear from
\refc{eqfinitesize} that the main contribution comes from the direction that enables the shortest handle.
Clearly this is the direction parallel to the spatial direction of the Wilson loop where $L'=L-R$.
The main problem thus arises for Wilson loops with large values of $R$.
For the case of our study, $2+1$ dimensional $SU(2)$ with $\beta=5.0$, a test for finite
volume effects with volumes of $24^3$ and $48^3$, excited states up to
$n=3$ and $R\leq12$ already exists \cite{BBPM}.
Since no finite volume effects were visible in this test, we can conclude that $a(L')$ is at most a
factor of $\Ord(1)$ and that we do not expect to suffer from any finite volume effects up to energy levels of $n=3$ at least.

\subsubsection{Control of the fits}
\label{chfit}

\begin{figure*}[t]
\centering
\begin{minipage}[c]{0.49\textwidth}
\centering
\includegraphics[angle=-90, width=\textwidth]{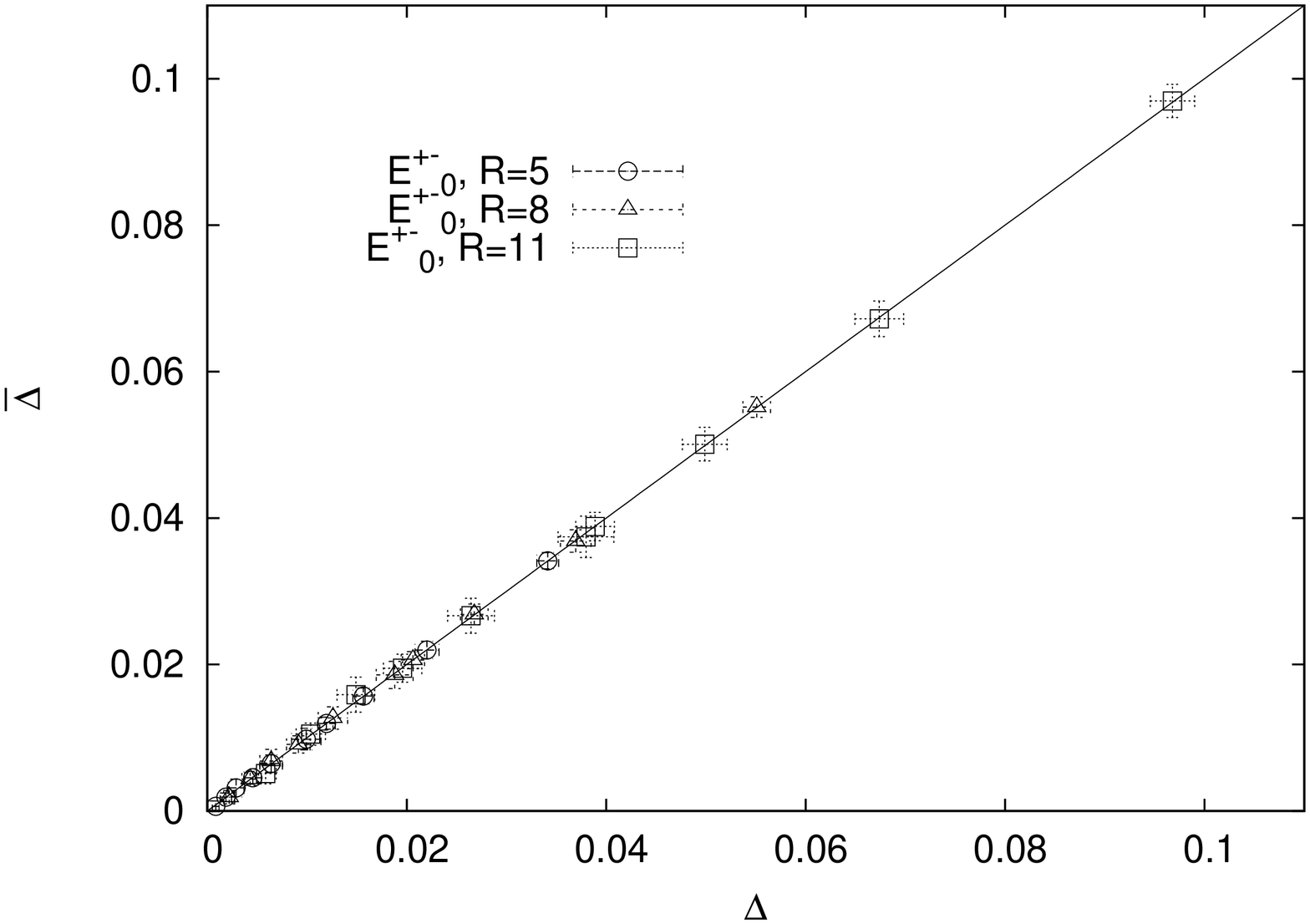}
\end{minipage}
\begin{minipage}[c]{0.49\textwidth}
\centering
\includegraphics[angle=-90, width=\textwidth]{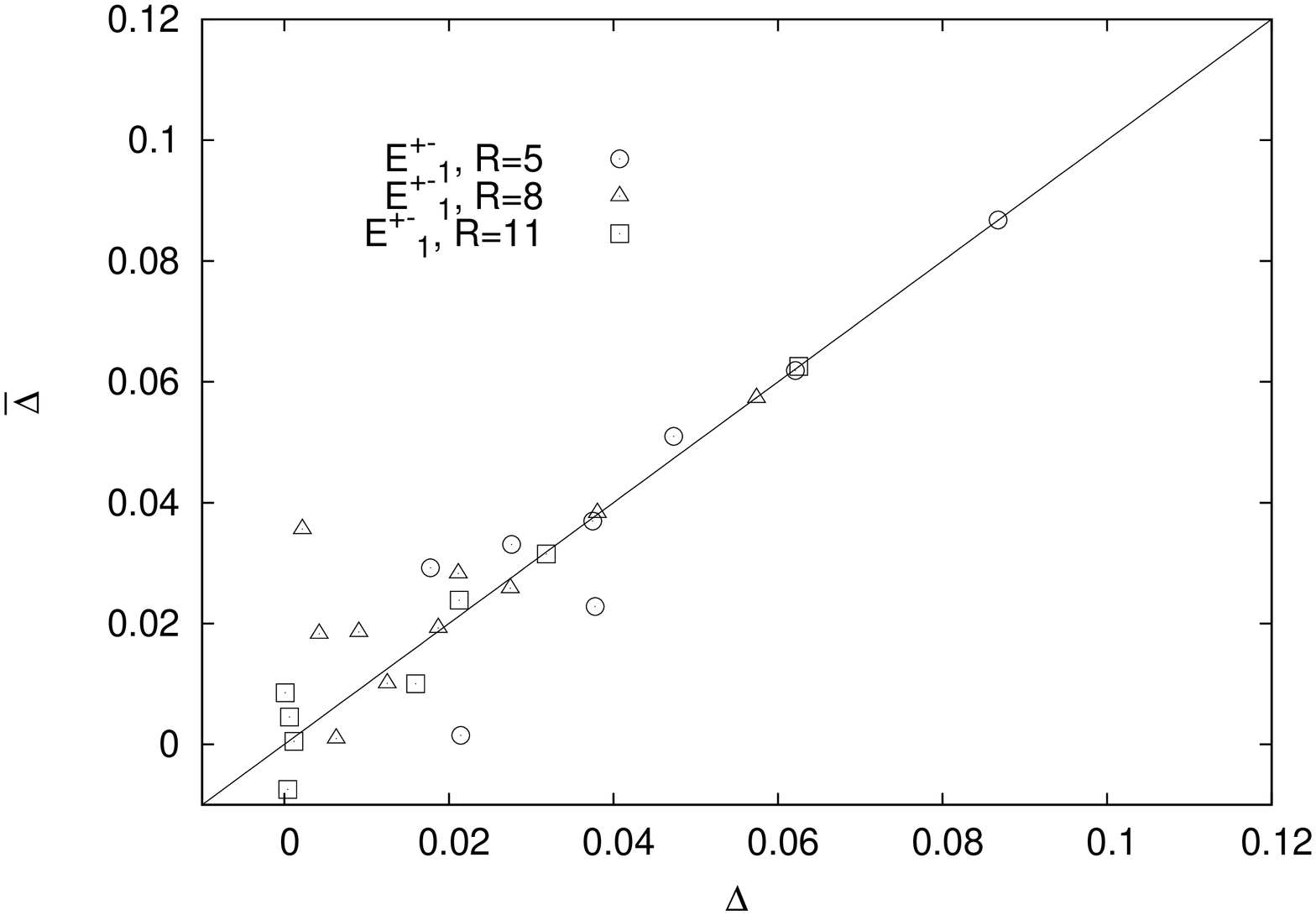}
\end{minipage}
\caption{Cross check of the fits for the corrected energies \refc{eqfit1} and energy differences \refc{eqfit2}: {\bf Left:} Fits for $E^{+-}_0$, $R=5,8$ and 11. These fits are expected to give reliable results. {\bf Right:} Fits for $E^{+-}_1$, $R=5,8$ and 11. These fits are regarded to be unreliable, since the points are clearly not following the $\Delta=\bar{\Delta}$-line. We have not plotted the error bars in the right fit, because they are large and confuse the picture.}
\label{fig2}
\end{figure*}

In the analysis a lot of fitting is done in order to extract the energies and energy differences
in the limit $T\to\infty$ and to take the contaminations from excited states into account.
For the control of systematic effects it is thus crucial to control these fits
in several ways, since these are the biggest source for systematic errors in the final data.

Even though the fits \refc{eqasympt} and \refc{eqasympt2} include only a linear fit function
it might be that already these fits give biased results, since the more important points
for the asymptotic behavior, the points from loops with large $T$, come with bigger statistical
errors. This might lead to the problem that the fit is completely determined by the values at small $T$.
As a check we plotted the data along with the resulting lines and checked the deviations
from the line of the points at large $T$. Where the deviations were not at the percent level and
below the statistical errors, we discarded the fit.

The second type of fits, eq. \refc{eqfit1} and \refc{eqfit2}, are much harder to control,
since these fits include nonlinear functions of two arguments. In addition the $\chi^2/d.o.f.$
of these fits is usually very small and does not provide much information
on the goodness of the fits.
We thus apply two checks that were already discussed in \cite{BBPM}.
We expect $\alpha$ to be smaller than the ratio of the degeneracies between the next excited state
and the state considered and $\delta$ to be of the order of the energy gap between the two.
Wherever this criterion was not fulfilled we did not use the fits.
The second check is a comparison between the expected corrections
\begin{equation}
\label{eqfitcontr}
\Delta= \frac{1}{T_{b}-T_{a}} \: \alpha^{CP}_{n}(R) \: e^{ - \delta^{CP}_{n}(R) \: T_{a} } \: 
\left( 1 - e^{ - \delta^{CP}_{n}(R) \: ( T_{b} - T_{a} ) } \right) \; ,
\end{equation}
obtained with averaged parameters $\alpha^{CP}_n$ and $\delta^{CP}_n$, and the difference
\begin{equation}
\label{eqfitdontr2}
\bar{\Delta} = E^{CP}_n(R) + \frac{1}{T_{b}-T_{a}} \: \ln \left[ \frac { \lambda^{CP}_{n} ( R , T_{b} ) } { \lambda^{CP}_{n} ( R , T_{a} ) } \right] \; ,
\end{equation}
for each $R$ and all combinations of $T_a$ and $T_b$.
We expect to obtain $\Delta=\bar{\Delta}$ for each $R$.
The plot of $\bar{\Delta}$ against $\Delta$ for three examples where we expect the fits to work are shown
on the left of figure \ref{fig2}, while on the right we show three examples
where the fits are regarded to be unreliable.
Similar checks were performed for the energy differences, too \cite{BBPM}.

\section{Results of the simulations}
\label{simresults}

\begin{figure}[t]
\centering
\includegraphics[angle=-90, width=.49\textwidth]{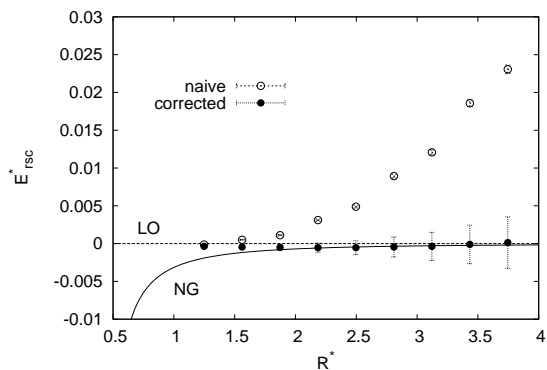}
\caption{Results for the ground state rescaled such that $E_0^{*,LO}\equiv0$. The lines are the LO and NG predictions as defined by \refc{eqarvis} in dimensionless quantities, eq. \refc{dimensquant}.}
\label{fig3}
\end{figure}

\begin{table*}[t]
\centering
\small
\begin{tabular}{|c|ll|ll|ll|ll|}
\hline
\multicolumn{1}{|c|}{} & \multicolumn{2}{c|}{$E^{++}_{0}=E_0$} & \multicolumn{2}{c|}{$E^{+-}_{0}=E_1$} & \multicolumn{2}{c|}{$E^{++}_{1}=E_{2,1}$} & \multicolumn{2}{c|}{$E^{--}_{0}=E_{2,2}$} \\
\hline
\multicolumn{1}{|c|}{$R$} & \multicolumn{1}{c|}{na\"ive} & \multicolumn{1}{c|}{corr} & \multicolumn{1}{c|}{na\"ive} & \multicolumn{1}{c|}{corr} & \multicolumn{1}{c|}{na\"ive} & \multicolumn{1}{c|}{corr} & \multicolumn{1}{c|}{na\"ive} & \multicolumn{1}{c|}{corr} \\
\hline
\hline
4 & 0.57281(3) & 0.5716(1) & 1.1278(2) & 1.118(1) & 1.452(2) & 1.35(6) & 1.5056(9) & \multicolumn{1}{c|}{---} \\
5 & 0.67623(4) & 0.6756(2) & 1.1643(2) & 1.152(1) & 1.491(1) & 1.41(2) & 1.5110(8) & 1.477(12) \\
6 & 0.77831(5) & 0.7775(2) & 1.2115(2) & 1.198(1) & 1.516(1) & 1.45(2) & 1.5296(8) & 1.511(7) \\
7 & 0.87969(7) & 0.8781(3) & 1.2735(2) & 1.255(1) & 1.581(1) & 1.51(1) & 1.5674(7) & 1.549(6) \\
8 & 0.98000(8) & 0.9779(4) & 1.3387(2) & 1.318(2) & 1.648(1) & 1.56(1) & 1.6137(7) & 1.595(5) \\
9 & 1.08047(9) & 1.0772(5) & 1.4159(2) & 1.387(2) & 1.731(2) & 1.62(1) & 1.6685(7) & 1.648(5) \\
10 & 1.18005(11) & 1.1761(6) & 1.4899(2) & 1.460(2) & 1.846(2) & 1.69(2) & 1.7296(8) & 1.709(4) \\
11 & 1.28020(13) & 1.2749(8) & 1.5766(2) & 1.537(2) & 1.940(3) & 1.76(2) & 1.7946(9) & 1.774(4) \\
12 & 1.37937(15) & 1.3734(9) & 1.6559(3) & 1.616(3) & 2.046(4) & 1.84(4) & 1.8644(9) & 1.843(3) \\
\hline
\end{tabular}
\normalsize
\caption{Results for the total energies in lattice units. In addition to the corrected data, which are our final results, we also list the na\"ive results. They give upper bounds for future simulations and illustrate the effect of contaminations from excited states. Note that also the na\"ive energies are already asymptotic results for $T\to\infty$, neglecting the contaminations.}
\label{tab2}
\end{table*}

\begin{table*}[t]
\centering
\small
\begin{tabular}{|c|ll|ll|ll|ll|}
\hline
\multicolumn{1}{|c|}{} & \multicolumn{2}{c|}{$E^{+-}_{0}-E^{++}_0$} & \multicolumn{2}{c|}{$E^{++}_{1}-E^{++}_0$} & \multicolumn{2}{c|}{$E^{--}_{0}-E^{++}_0$} & \multicolumn{2}{c|}{$E^{++}_{1}-E^{+-}_0$} \\
\multicolumn{1}{|c|}{} & \multicolumn{2}{c|}{$=\Delta E_{10}$} & \multicolumn{2}{c|}{$=\Delta E_{20}$} & \multicolumn{2}{c|}{$=\Delta E_{20}$} & \multicolumn{2}{c|}{$=\Delta E_{21}$} \\
\hline
\multicolumn{1}{|c|}{$R$} & \multicolumn{1}{c|}{na\"ive} & \multicolumn{1}{c|}{corr} & \multicolumn{1}{c|}{na\"ive} & \multicolumn{1}{c|}{corr} & \multicolumn{1}{c|}{na\"ive} & \multicolumn{1}{c|}{corr} & \multicolumn{1}{c|}{na\"ive} & \multicolumn{1}{c|}{corr} \\
\hline
\hline
4 & 0.5558(2) & 0.547(1) & \multicolumn{1}{c}{---} & \multicolumn{1}{c|}{---} & 0.9334(9) &  \multicolumn{1}{c|}{---} & \multicolumn{1}{c}{---} & \multicolumn{1}{c|}{---} \\
5 & 0.4879(2) & 0.476(2) & 0.814(1) & 0.734(18) & 0.8341(8) & 0.800(12) & 0.319(1) & \multicolumn{1}{c|}{---} \\
6 & 0.4330(2) & 0.421(2) & 0.736(1) & 0.672(14) & 0.7506(8) & 0.733(7) & 0.296(1) & 0.25(2) \\
7 & 0.3936(2) & 0.377(2) & 0.699(2) & 0.628(11) & 0.6867(7) & 0.671(6) & 0.296(2) & 0.25(1) \\
8 & 0.3584(2) & 0.340(2) & 0.666(1) & 0.585(9) & 0.6325(7) & 0.617(5) & 0.299(1) & 0.24(1) \\
9 & 0.3351(2) & 0.310(2) & 0.648(2) & 0.539(11) & 0.5684(8) & 0.571(5) & 0.302(2) & 0.23(1) \\
10 & 0.3094(2) & 0.283(2) & \multicolumn{1}{c}{---} & 0.513(16) & 0.5477(8) & 0.533(4) & \multicolumn{1}{c}{---} & 0.23(2) \\
11 & 0.2956(2) & 0.261(3) & \multicolumn{1}{c}{---} & 0.483(19) & 0.5122(9) & 0.499(3) & \multicolumn{1}{c}{---} & 0.22(2) \\
12 & 0.2757(2) & 0.242(3) & \multicolumn{1}{c}{---} & 0.467(35) & 0.4828(9) & 0.469(3) & \multicolumn{1}{c}{---} & 0.22(4) \\
\hline
\end{tabular}
\normalsize
\caption{Results for the energy differences in lattice units. As in table \ref{tab2} we list corrected and na\"ive results. For some differences we were able to obtain corrected but no na\"ive results. This is due to the fact that the contaminations from excited states are large and the data points for the fit \refc{eqasympt2} do not lie on a straight line, whereas the fit \refc{eqfit2} still works well.}
\label{tab3}
\end{table*}

We list all results, na\"ive and corrected in tables \ref{tab2} and \ref{tab3},
to enable other groups to use the data for comparisons.
In this section we compare our data to the full NG energy spectrum \refc{eqarvis}
and its truncations in $1/R^2$
to leading order (LO), next-to leading order (NLO) and next-to-next-to leading order (NNLO).
The comparison to corrections from boundary terms is postponed to the next section.

In order to compare to the data
we have to extract the string tension $\sigma$ and fix an unphysical constant $V_0$.
Since we expect corrections
to appear with respect to the full NG prediction (as concluded e.g. in \cite{teper2} and \cite{r4cor}),
we fit the ground state data to the form
\begin{equation}
\label{eqsigfit}
V(R) = \sigma \: R \: \sqrt{ 1 - \frac{\pi}{12\:\sigma\:R^{2}} } + V_{0} \; .
\end{equation}
Since we expect string like behavior to set in at $\sim 2\:r_0$ \cite{BBPM,PundD1,PundD2},
we omit the first three points in the fit and obtain
\begin{equation}
\label{eqsigmares}
\sigma = 0.0975(2) \quad \textnormal{and} \quad V_0 = 0.2148(7) \; .
\end{equation}

This result for $\sigma$ is consistent with the value from \cite{PundD1,PundD2} within the error bars.
The error bars are much larger in our case,
which is no surprise since our aim is not to obtain the ground state energy and the
string tension with high accuracy and our setup is not tuned to reduce this error in particular.
From now on we are going to use the dimensionless quantities
\begin{equation}
\label{dimensquant}
R^{*}\equiv \sqrt{\sigma}\:R \quad \textnormal{and} \quad E^{*}\equiv \left(E-V_0\right)/\sqrt{\sigma}
\end{equation}
for plotting and comparisons to the predictions.

\begin{figure*}[t]
\centering
\begin{minipage}[c]{0.49\textwidth}
\centering
\includegraphics[angle=-90, width=\textwidth]{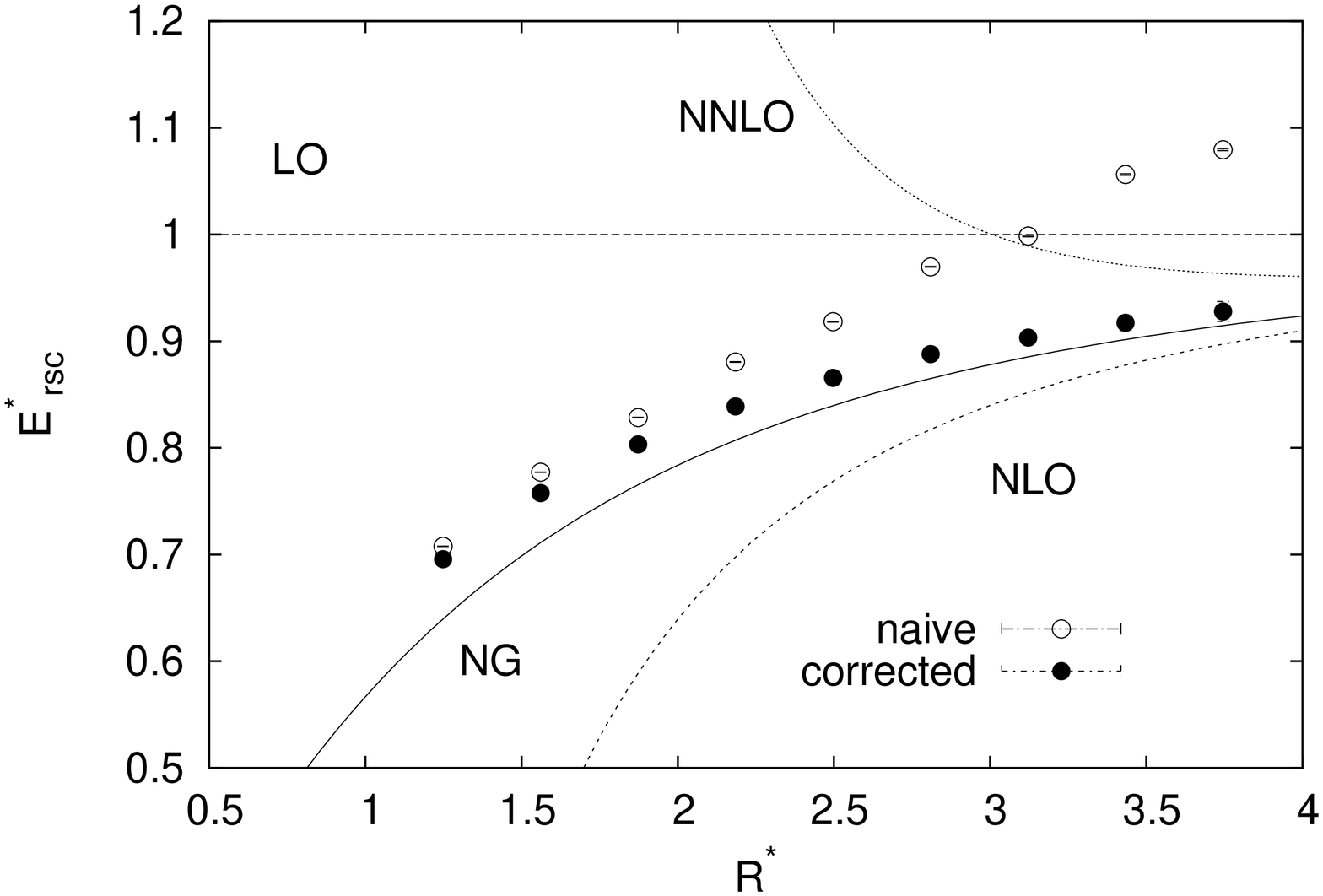}
\newline
\includegraphics[angle=-90, width=\textwidth]{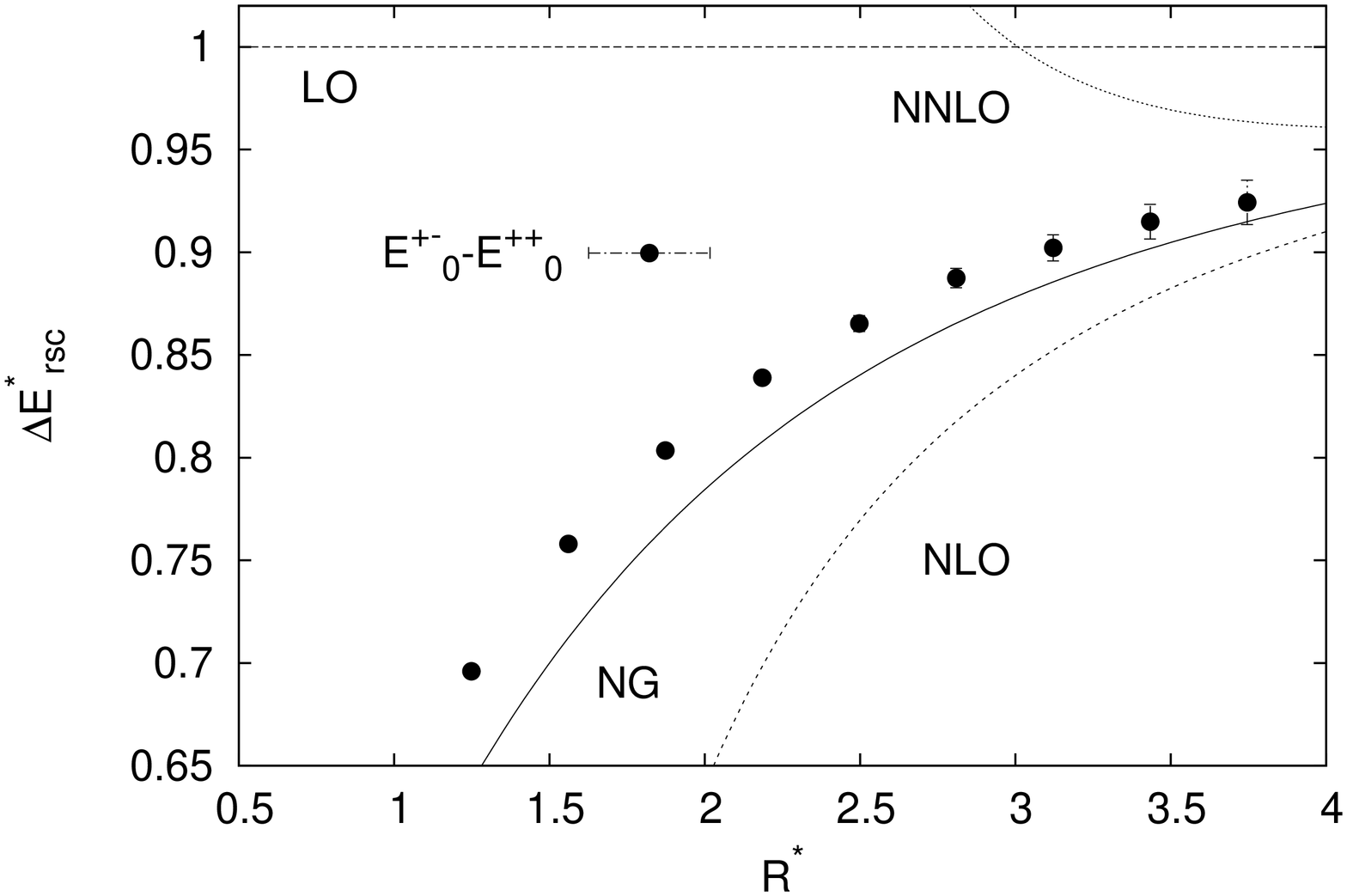}
\end{minipage}
\begin{minipage}[c]{0.49\textwidth}
\centering
\includegraphics[angle=-90, width=\textwidth]{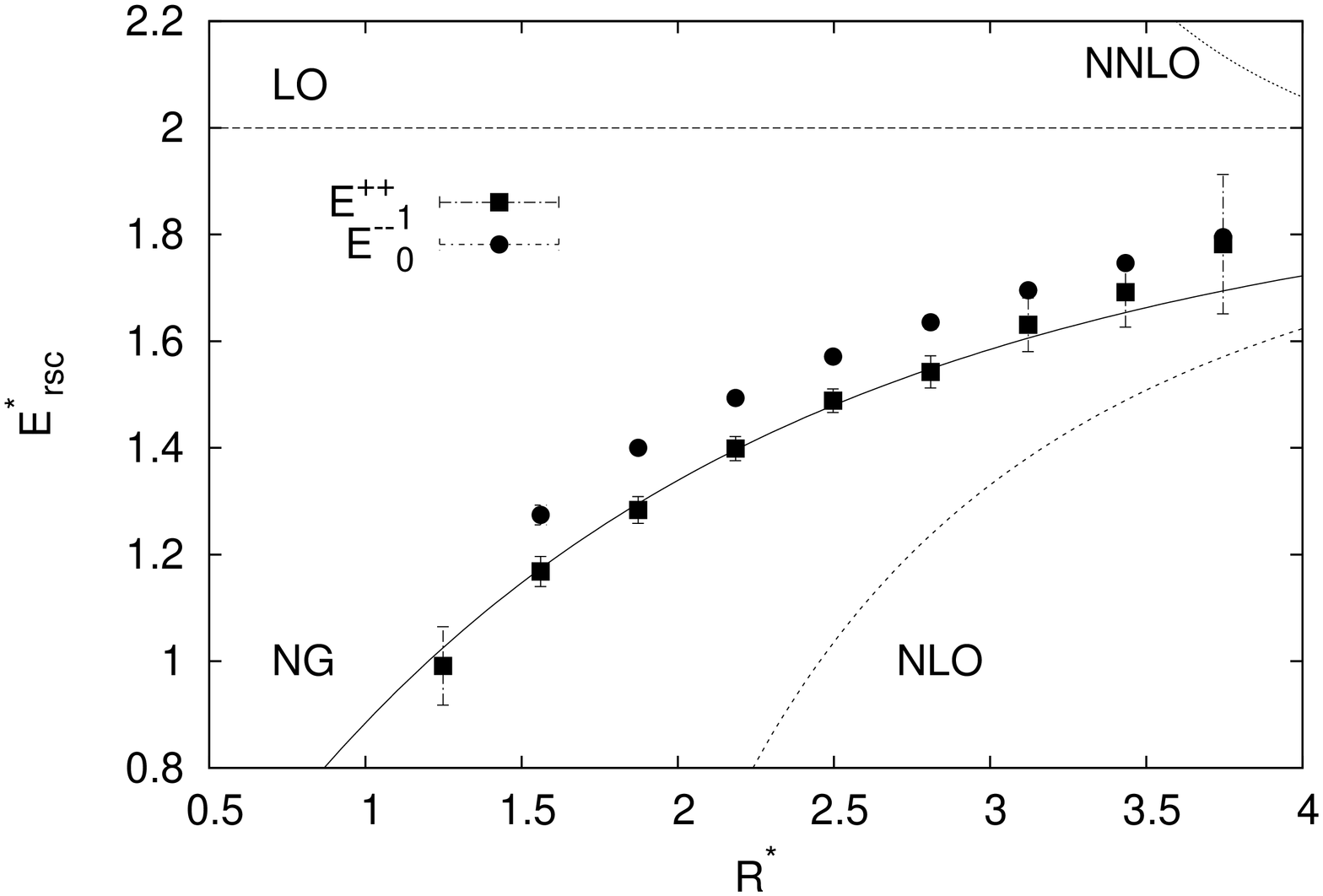}
\newline
\includegraphics[angle=-90, width=\textwidth]{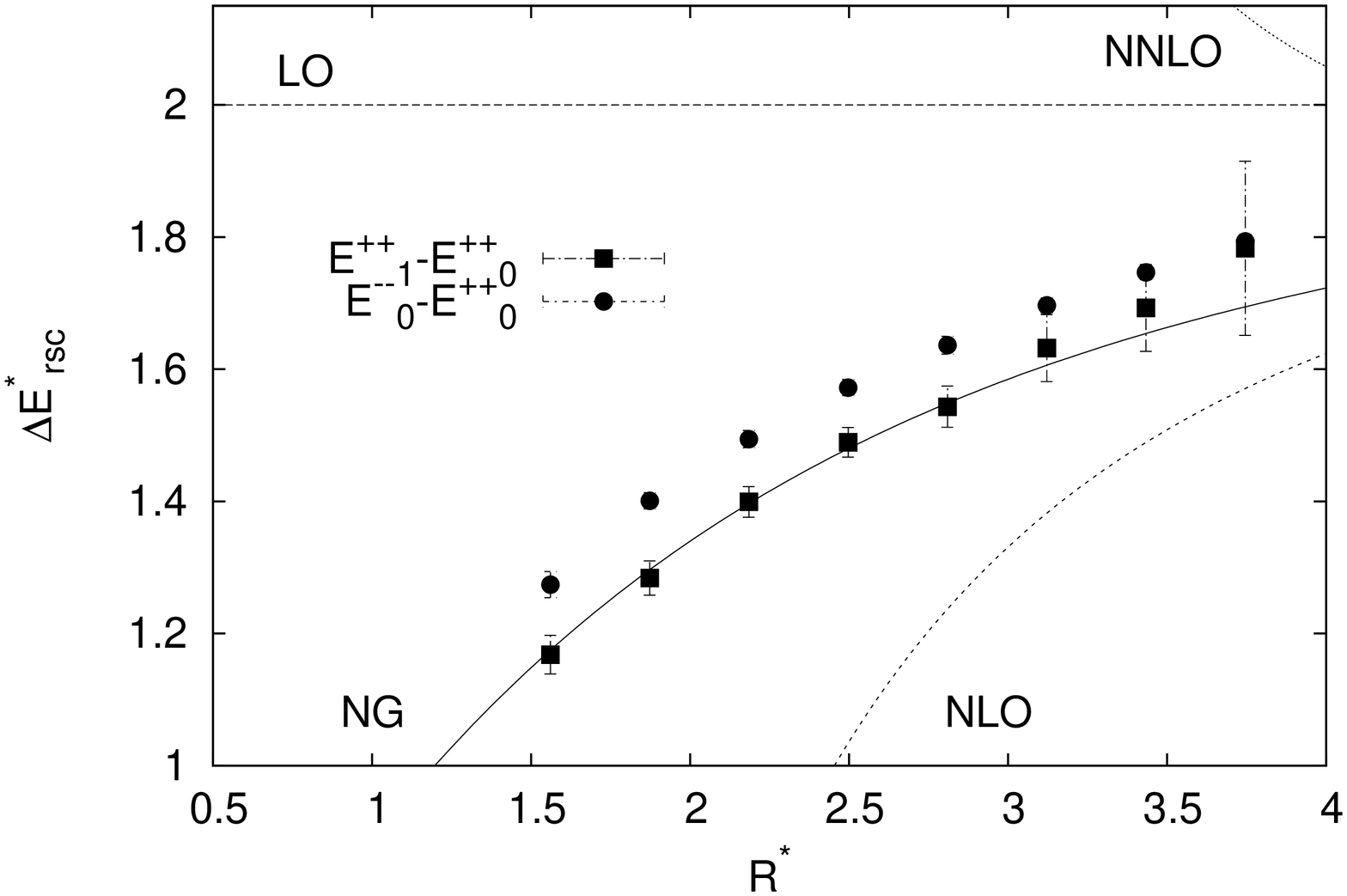}
\end{minipage}
\caption{{\bf Top:} Results for the first (left) and second excited state (right), rescaled such that $E_n^{*,LO}\equiv n$. {\bf Bottom:} Results for the energy differences corresponding to $\Delta E_{10}$ (left) and $\Delta E_{20}$ (right) in the free theory. The data is rescaled such that $\Delta E^{*,LO}_{nm}=n-m$.}
\label{fig4}
\end{figure*}

We plot the na\"ive and corrected results for the ground state in figure \ref{fig3}.
In the plots we have rescaled the energies such that $E^{*,LO}_n\equiv n$, i.e.
\begin{equation}
\label{eq-rescale}
E^*_{n,rsc}(R^*) = \left( E^{*}_n(R^*) - R^* \right) \: \frac{R^*}{\pi} + \frac{1}{24} \; ,
\end{equation}
to provide visibility of small effects.
We see a clear splitting between the na\"ive and the corrected results,
which increases when we go to larger values of $R^*$.
The effect becomes even more severe when we go to the excited
states, as shown for energy level $E_1$ in figure \ref{fig4} (top left).
We can also see the approximate linear asymptotic behavior of the rescaled na\"ive energies
when going to large $R$ values, which was seen in \cite{Juge1,Juge2,Juge3}, too.
The asymptotic behavior of the corrected energies is different 
and we conclude that the behavior of the na\"ive energies is due to contributions from excited states
and does not reflect the physical asymptotic behavior of the energies for large $R$.
When we look at the predictions for the energy gaps from \refc{eqarvis} and its truncations
it is easy to see why the excited state contamination is enhanced.
The energy gaps to the excited states decrease roughly like $1/R$,
thus the damping with $T$ is reduced in eq. \refc{eigst2} when going to larger $R$
and more excited states contribute to the contamination.
In the following we therefore only use corrected results for the discussions.

The results for $E^{++}_1$ and $E^{--}_0$, belonging to the second excited state,
are shown in figure \ref{fig4} (top right).
It is remarkable that with the new method and 5 different
temporal extents we were able to reduce the errors for the corrected values of the ground state
in the $(+,-)$-channel
by a factor of more than two and to obtain corrected results for the groundstate in the $(-,-)$-channel
with an accuracy better than that of the results for the ground state
of the $(+,-)$-channel in \cite{BBPM}.
The error bars of the first excited state in the $(+,+)$-channel are almost an order
of magnitude bigger than the
error bars from the ground state of the $(-,-)$-channel, but nevertheless we see a $2-4$ sigma splitting
up to $R^*\approx3$. At that point the error of $E^{++}_1$ becomes to large to distinguish between
the two.

The energy differences are listed in table \ref{tab3}.
They have the advantage
that the comparison to the predictions does not involve the unknown constant $V_0$ from eq.~\refc{eqsigfit}
and that they are more sensitive to subleading properties of the flux tube.
In figure \ref{fig4} we show the difference $E^{+-}_0-E^{++}_0$ (left bottom),
corresponding to the free difference $\Delta E_{10}$,
and the differences $E^{++}_1-E^{++}_0$ and $E^{--}_0-E^{++}_0$ (right bottom),
corresponding to the free difference $\Delta E_{20}$.
We see a similar error reduction as for the absolute energy values compared to \cite{BBPM}.

The qualitative agreement between the data and the NG predictions is
remarkable down to very small values of $R^*$, where we do not expect the flux tube to have
a string like shape and where the expansion of the string action in $1/R^*$
is no longer expected to converge.
These findings are in full agreement with other simulations, e.g. \cite{Pushan2a,Pushan2b,ownPos,BBPM}.
In addition they are consistent with the results from \cite{LW-calc,Juge1,Juge2,Juge3} for smaller values
of $R^*$ and with the results obtained for closed strings \cite{teper1,teper2}.
Even though the data follows the NG predictions qualitatively,
we see a significant deviation for the absolute energies
$E^{+-}_0$ and $E^{--}_0$, as well as for the corresponding differences to the ground state,
even for our largest values of $R^*$. We are going to compare these deviations to the
boundary corrections in the next section.
The $E^{++}_1$ state is fully consistent with the NG prediction inside the error bars
for all values of $R^*$, which is also true for the corresponding difference $E^{++}_1-E^{++}_0$.

\section{Discussion of $1/R^4$ corrections}
\label{r4cor-ch}

To compare the boundary corrections to the NG predictions with the data, we have to extract the
unknown coefficient $b_2$.
There is a hint from theory that $b_2$ is negative and nonvanishing,
coming from a computation of $b_2$ for the case of a
string ending on two d-branes in confining gauge theories with a weakly curved holographic dual
\cite{ahar2}.
Corrections appears at all energy levels
and we demand that $b_2$ is consistent with all data above the critical distance as defined in section
\ref{corr}. We therefore use the ground state as well as the first excited state
to extract $b_2$ from a simultaneous fit.
We also have to account for the possibility that the values for $\sigma$ and $V_0$
are changed due to the ground state correction.
We thus use them again as free parameters
denoted as $\sigma'$ and $V'_0$ to distinguish them from the values obtained in the last section.

\begin{table*}[t]
\centering
\small
\begin{tabular}{c|cc|c|cc|cc|cc}
\hline
fit & $\sigma'$ & $V'_0$ & $b_2$ & $\gamma_0$ & $\gamma_1$ & $\eta_0$ & $\eta_1$ & $\chi^2/d.o.f.$ \\
\hline
\hline
1 & 0.0974(1) & 0.2151(4) & -0.30(4) & --- & --- & --- & --- & 0.79 \\
\hline
2 & 0.0975(2) & 0.2145(7) & -0.61(19) & 2(3) & -2(1)$\cdot10^3$ & --- & --- & 0.04 \\
\hline
3 & 0.0975(2) & 0.2145(6) & -0.52(14) & --- & --- & 14(11) & -11(6)$\cdot10^3$ & 0.05 \\
\hline
4 & 0.0975(2) & 0.2146(7) & --- & 9(3) &  12(3)$\cdot10^3$ & --- & -7(2)$\cdot10^4$ & 0.22 \\
\hline
\end{tabular}
\normalsize
\caption{Results for the combined fit of ground state and first excited state data to the form \refc{eqcorfit}. The different fits are described in the text.}
\label{tab4}
\end{table*}

It is known that, when fitting to a polynomial, the coefficient of the highest power of the polynomial in
the fit receives a summed contribution from possible higher order terms.
This spoils the reliability of the highest coefficient of the fit polynomial.
$b_2$ suffers from the same problem, since the fit function is a nonlinear function in $\sigma'$
combined with a polynomial correction.
To improve the reliability of the obtained value for $b_2$ we
include a correction term of higher order to each energy level.
There are in general two possibilities at which order the next correction term might appear.
At $\Ord(R^{-6})$ if there are additional boundary terms in the effective action or at $\Ord(R^{-7})$
if the next correction term is a regular (non-boundary) correction.
To account for these possibilities we parametrise our fit functions for a simultaneous
fit to the ground state and the first excited state as:
\begin{equation}
\label{eqcorfit}
\begin{array}{rl}
\displaystyle E_0(R) = & \displaystyle \sigma' \: R \: \sqrt{ 1 - \frac{\pi}{12\:\sigma'\:R^{2}} }
 - b_2 \: \frac{\pi^{3}}{60} \: \frac{1}{R^4} + \gamma_0 \: \frac{1}{R^6} + \eta_0 \: \frac{1}{R^7} + V'_{0} \vspace*{2mm} \\
\displaystyle E_1(R) = & \displaystyle \sigma' \: R \: \sqrt{ 1 + \frac{23\:\pi}{12\:\sigma'\:R^{2}} }
 - b_2 \: \frac{241\:\pi^{3}}{60} \: \frac{1}{R^4} + \gamma_1 \: \frac{1}{R^6} + \eta_1 \: \frac{1}{R^7} + V'_{0} \; ,
\end{array}
\end{equation}
To check the effect of the higher order terms we perform the following fits:
\begin{enumerate}
 \item Set $\gamma_i\equiv0$ and $\eta_i\equiv0$, use $b_2$ as a fit parameter.
 \item Set $\eta_i\equiv0$, use $b_2$ and $\gamma_i$ as fit parameters.
 \item Set $\gamma_i\equiv0$, use $b_2$ and $\eta_i$ as fit parameters.
 \item Set $b_2\equiv0$ and $\eta_0\equiv0$, use $\gamma_i$ and $\eta_1$ as fit parameters.
\end{enumerate}
In all cases $\sigma'$ and $V'_0$ are free parameters as well.
The last fit accounts for the possibility that the coefficient $b_2$ vanishes identically,
which is not yet ruled out completely by theory and thus remains a possibility.
In the fits we include all points except the one with smallest $R$ for the ground state,
and all points with $R\geq7$ for the first excited state,
since we expect the data to be consistent with the effective string theory even for smaller values of
$R$ when the corrections are included.
We exclude the data for the second excited state for the following reasons:
Only the largest values of $R^*$ are above the critical length below which
the expansion of the energy ceases to be convergent for the second excited state (see eq. \refc{eqradconv}) and one has to include four more fit parameters
$\gamma_{2,i}$ and $\eta_{2,i}$ to the fits.
Therefore the inclusion of the second excited state does not improve the
information and the accuracy for $b_2$.
We do not account for the possibility of correction terms below $\Ord(R^{-4})$,
since we compare to the effective theory where these
terms are explicitly ruled out.

\begin{figure*}[t]
\centering
\begin{minipage}[c]{0.49\textwidth}
\centering
\includegraphics[angle=-90, width=\textwidth]{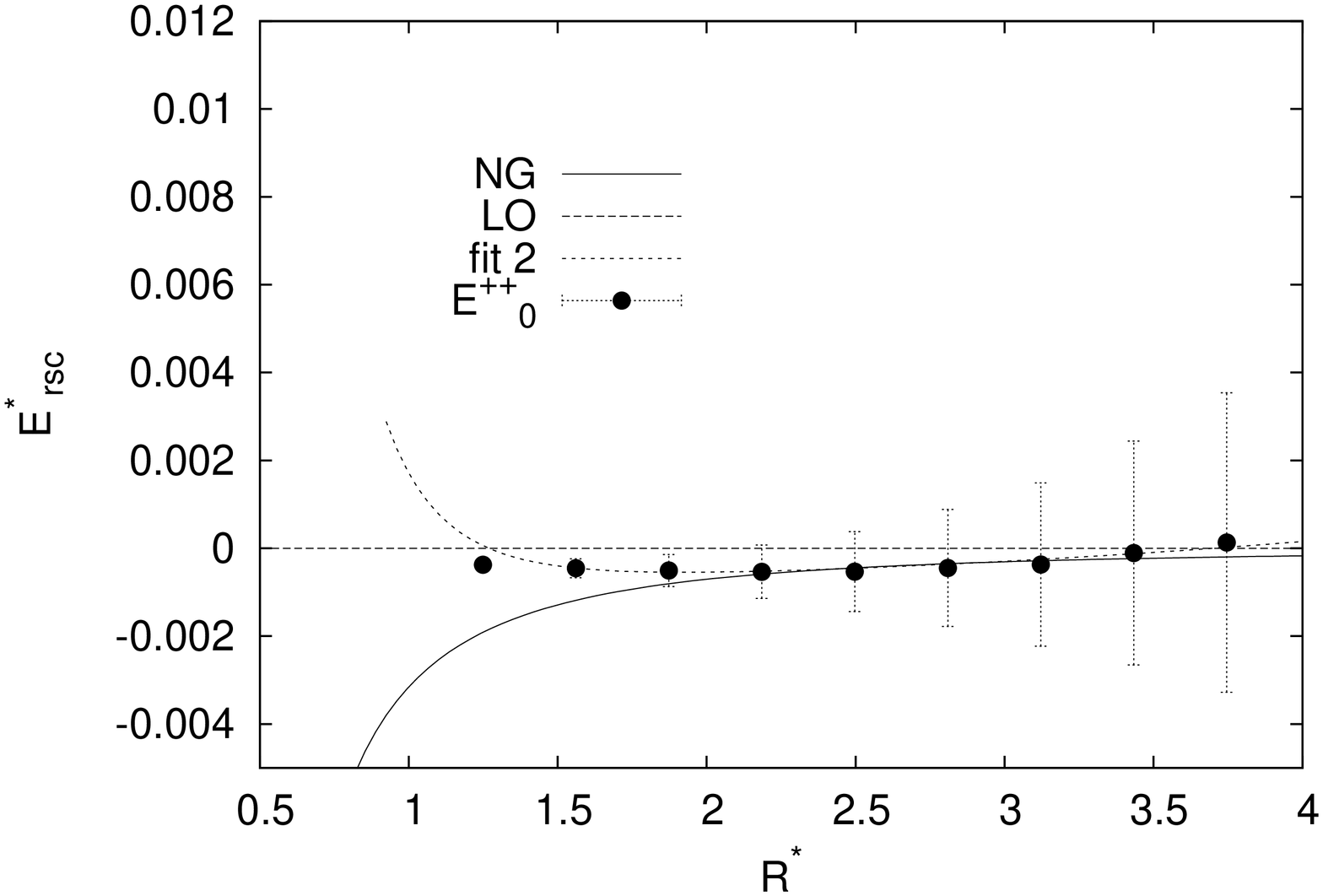}
\end{minipage}
\begin{minipage}[c]{0.49\textwidth}
\centering
\includegraphics[angle=-90, width=\textwidth]{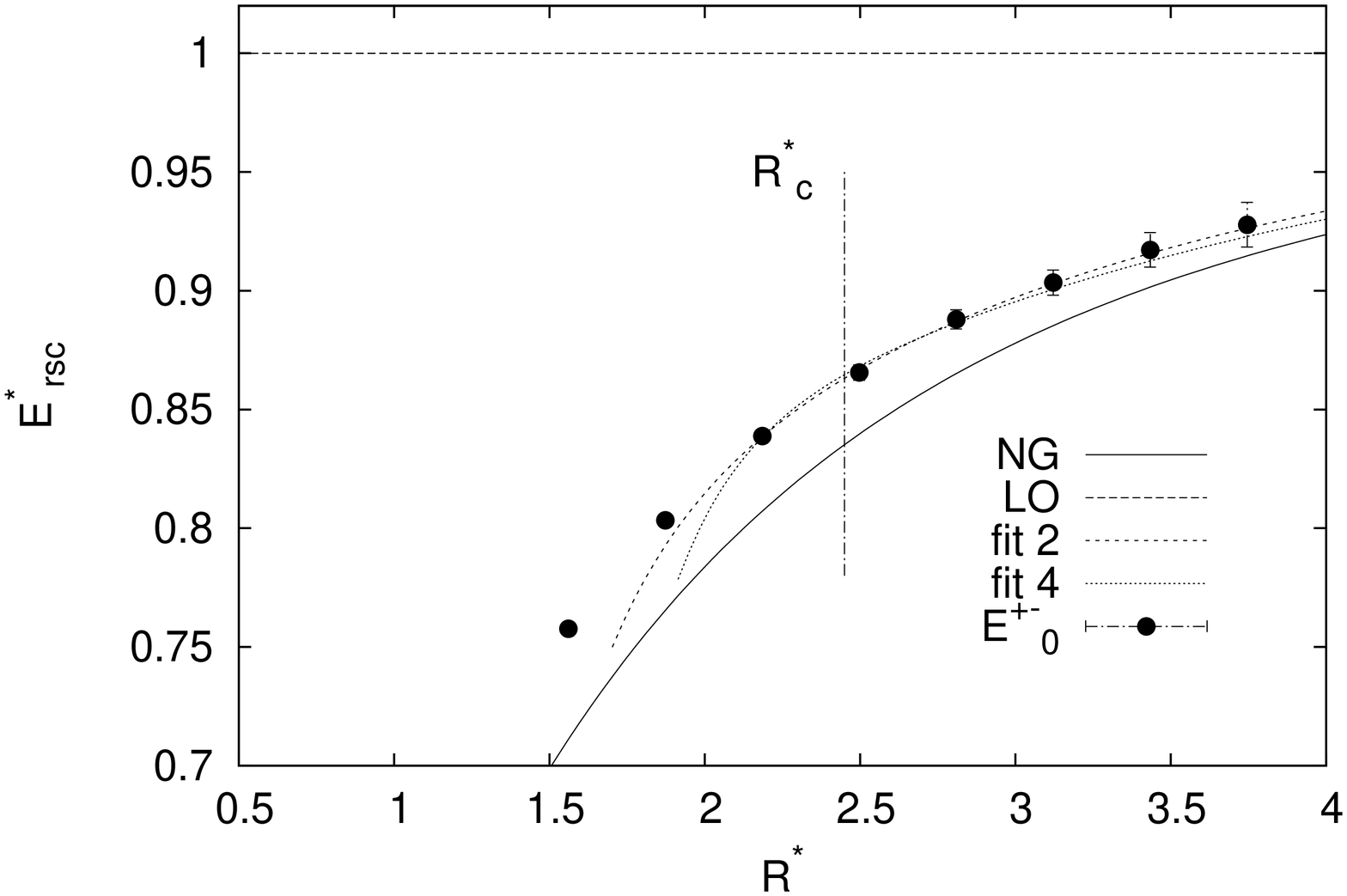}
\end{minipage}
\caption{Comparison between fit 2 from table \ref{tab4} and the data for the groundstate (left) and the first excited state (right). The data is rescaled as in the previous figures. The line marked with $R^*_c$ is the line below which the expansion of the corresponding energy level in $1/R^*$ ceases to be convergent.}
\label{fig5}
\end{figure*}

We list the results of the fits in table \ref{tab4}. In all cases we see good agreement between
$\sigma'$ and $V'_0$ and $\sigma$ and $V_0$
obtained in the previous section. This shows that the extraction of $\sigma$ from the
ground state is not sensitive to $1/R^4$ corrections within the accuracy for the
ground state data of our simulations.
The result for $b_2$ varies between fits $1-3$, but is small and negative in all cases.
We expect the results for $b_2$ from the fits 2 and 3,
including also the higher order terms, to be unbiased.
In that case $\gamma_0$ or $\eta_0$ is practically consistent with zero,
which is not surprising, since we expect the data for the ground state not to be accurate enough
to make higher order corrections to a $1/R^4$ correction visible.
$\gamma_1$ and $\eta_1$ are nonzero but still small compared to the NG coefficient at $\Ord(R^{-7})$.
Fit 4 seems to work as well but a relatively large nonvanishing boundary term
at $\Ord(R^{-6})$ is needed for the fit to work accurately.
Nevertheless the corresponding $\chi^2/d.o.f.$ is a magnitude bigger
than for the fits 2 and 3 (even though it is still below one).
Assuming that $b_2$ does not vanish identically we find the data to be consistent with
\begin{equation}
 \label{resb2}
b_2=-0.5(2)(2) \; ,
\end{equation}
where the first error is statistical and the second reflects the systematical uncertainty,
estimated from the variation of $b_2$ between fits 1-3.

To compare to the data, we use fit 2 from table \ref{tab4} and
show the corresponding plots for the ground state and the first excited state in figure \ref{fig5}.
We use $\sigma$ and $V_0$ as before (instead of $\sigma'$ and $V'_0$)
to rescale the data points and show also the NG lines from the previous section
for comparison.
It is remarkable that we can accurately describe the corrections to two different energy states with
a predicted splitting and a single coefficient.
In addition we also show the resulting curve from fit 4 for the first excited state.
We see that it deviates a little bit more from the data than the curve obtained from fit 2,
but is well inside the error bars and describes the data as well.

\begin{figure}[t]
\centering
\includegraphics[angle=-90, width=.49\textwidth]{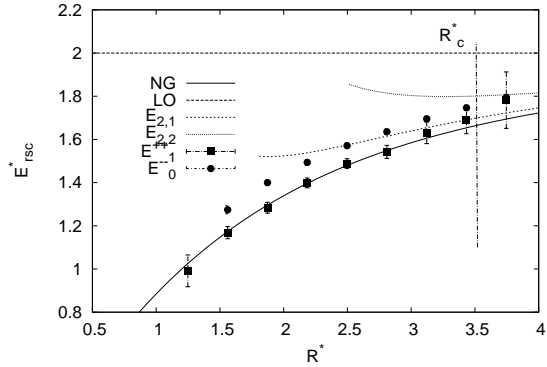}
\caption{Comparison between the data for the second excited state and the corrected energy levels $E_{2,1}=E^{NG}_2+\epsilon_{2,1}$ and $E_{2,2}=E^{NG}_2+\epsilon_{2,2}$ with $b_2$ obtained from fit 2. The data is rescaled as in the previous figures.}
\label{fig6}
\end{figure}

Having fixed $b_2$ through the fit (in this case fit 2),
we now get a prediction for the two
curves corresponding to the second excited state, where the degeneracy is lifted.
We show the corresponding curves together with the data in figure \ref{fig6}.
Additional terms of $\Ord(R^{-6})$ or higher are excluded here, since we do not know anything about
their coefficients and we expect the effects to be
of minor importance at the present level of accuracy of the data.
We see that the curve $E_{2,1}$ stays below the $E_{2,2}$ curve which is consistent with the
behavior of the data.
Nevertheless the data is closer to the NG curve when $R^*<3$,
which is well below $R^*_c$ for the second excited state.
Above $R^*_c$ the data for $E^{--}_0$ approaches the $E_{2,2}$ curve, but
it is not clear whether the behavior with increasing $R^*$ remains consistent with the resulting curve.
The data for $E^{++}_1$ is consistent with the $E_{2,1}$ curve already below $R^*_c$,
but the error bars above $R^*_c$ are to large to exclude the approach to any of the three curves.

\begin{figure*}[t]
\centering
\begin{minipage}[c]{0.49\textwidth}
\centering
\includegraphics[angle=-90, width=\textwidth]{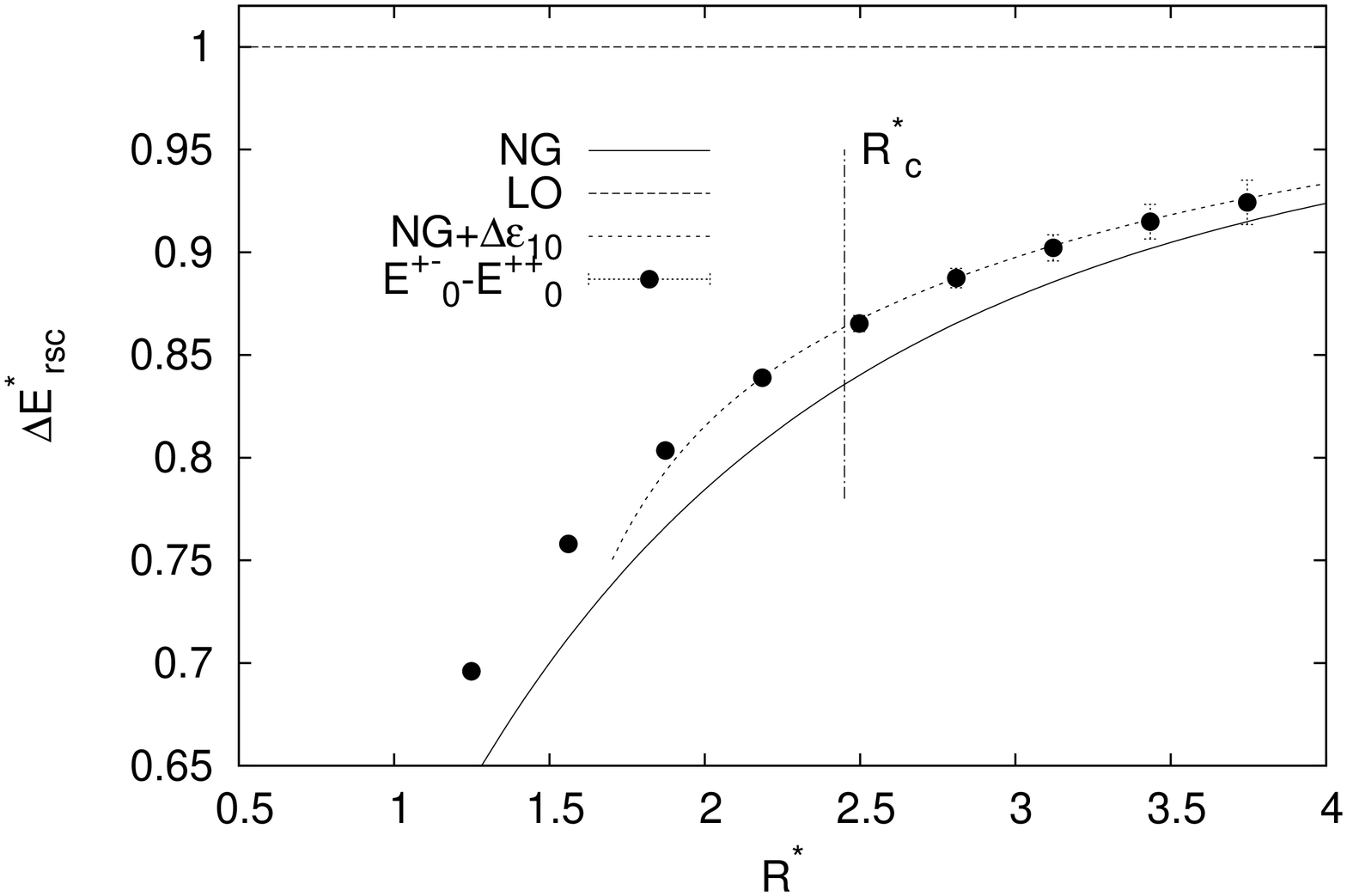}
\end{minipage}
\begin{minipage}[c]{0.49\textwidth}
\centering
\includegraphics[angle=-90, width=\textwidth]{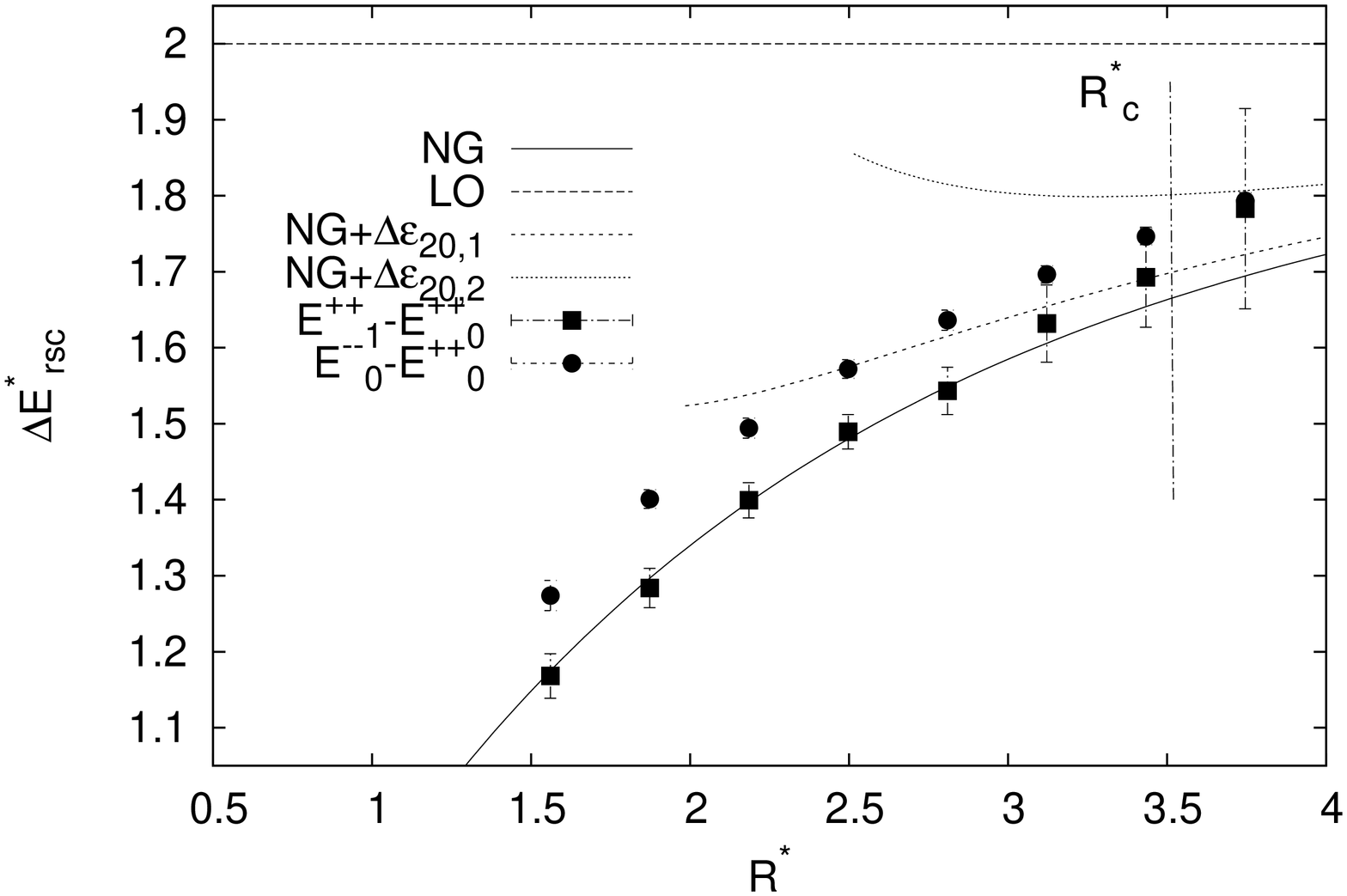}
\end{minipage}
\caption{Comparison between the data for the energy differences $\Delta E_{10}$ (left) and $\Delta E_{20}$ (right) and the predictions with $b_2$ obtained from fit 2. The data is rescaled as in the previous figures. We have used the abbreviations $\Delta \epsilon_{10}=\epsilon_1-\epsilon_0$, $\Delta \epsilon_{20,1}=\epsilon_{2,1}-\epsilon_0$ and $\Delta \epsilon_{20,2}=\epsilon_{2,2}-\epsilon_0$.}
\label{fig7}
\end{figure*}

The energy differences provide another check for the fits of this section.
We show the curves obtained with the parameters from the fit above
and the corresponding data in fig. \ref{fig7}.
Above $R^*_c$ we see perfect agreement between the $\Delta E_{10}$
curve including the correction and the data.
For $\Delta E_{20}$ the picture is similar as for the total energies
at $n=2$, the data is roughly consistent with the predictions.

We can employ yet another independent check and use the splitting between first excited
state and ground state directly to obtain $b_2$.
In that case we can keep $\sigma$ and $V_0$ fixed, which are almost completely determined
by the groundstate.
When we include an additional correction term at $\Ord(R^{-6})$ we obtain $b_2=-0.56(23)$ with
$\chi^2/d.o.f.=0.03$. This result is consistent with the results from the fits to the total energies.
All in all, together with the hint from the effective theory that $b_2$ should be nonvanishing,
we consider the possibility of a vanishing $1/R^4$ correction to be unlikely with our
present data. The data for ground state and first excited state
is well described by the predicted splitting of the effective theory
with a single coefficient $b_2$.
Also the behavior of the second excited state, which was not included in the fits,
suits to the predictions.

\section{Conclusions}

In this article we have looked at the energies of the excited states of the
flux tube between static quark and antiquark in
three-dimensional $SU(2)$ gauge theory, using a combination
of the algorithm from \cite{BBPM} and a variational method.
We studied $q\bar{q}$ separations between $1$ and $3\:r_0$ and energy levels up
to $n=2$.
We have discussed the main systematic effects entering the extraction of the excited states in detail,
explained under which conditions we consider the results to be unaffected by any of these effects
and included only those results in the analysis where systematics are under control.
The only remaining systematic effect which is not investigated is the lattice spacing dependence,
but results from \cite{Pushan2a,Pushan2b,ownPos,BBPM} suggests that the dependence on the lattice
spacing is very mild.
Using the correlation matrices in the $(C,P)$ channels and 5 different temporal extents,
we were able to further reduce the errors compared to \cite{BBPM} and to obtain
a clear signal for the second excited state.
In addition we were able to obtain results for the first excited state in the $(+,+)$ channel,
which is degenerate to the ground state in the $(+,-)$ channel in the free theory.
The effective string theory predicts a splitting of these formerly degenerate states and indeed
we observe such a splitting in our data.
In all cases we see the same qualitative agreement with the NG predictions as observed for
previous measurements in the three-dimensional case, e.g. \cite{Pushan2a,Pushan2b,ownPos,BBPM}.

Nevertheless, with enhanced accuracy and improved control over systematic effects we see a significant
deviation from the NG curves and compare it with the predictions from \cite{ahar2,r4cor}.
Our data is in good agreement with the predicted $1/R^4$ boundary correction to the
effective string theory, even though we cannot exclude
the possibility that boundary corrections do not appear until $\Ord(R^{-6})$.
If we assume that $b_2$ does not vanish identically we find evidence that $b_2$
is small and negative, $b_2=-0.5(2)(2)$, where the
first error is statistical and the second systematic.
The resulting curves are in good agreement with the data and it is remarkable that we can describe
the data accurately with the predicted splitting.
The negative sign of the result is consistent with the sign of the result for
confining gauge theories with a weakly curved holographic dual, applied to the case of
a string ending on two d-branes \cite{ahar2}.
This is the first time that significant deviations from NG energy levels could be observed
for open strings in three dimensions.
In the case of closed strings and in the finite temperature behavior of the string tension
analogous deviations were observed in \cite{teper2} and \cite{gliozzi}.

To confirm the findings of this paper it is desirable to check the consistency with earlier results,
as e.g. the results for the ground state from \cite{PundD1,PundD2},
and to perform simulations at additional lattice spacings.
One would also like to have a precise prediction for the coefficient $b_2$ to compare with,
which is hard to get if $b_2$ is not constrained by any symmetry.
In addition it would be interesting to see whether $b_2$ is universal
or whether it varies between different theories.
To clarify the picture one also has to push the simulations to bigger values of $R$,
in order to reach the region of a convergent expansion for the higher energy levels,
and to further increase the precision for the excited states.

\acknowledgments

The simulations were mainly done on the Linux cluster LC2 at the ZDV of the Johannes
Gutenberg-Universit\"at Mainz. A smaller part was done on ''Lilly'' at the Institut f\"ur Kernphysik.
I am thankful to the institutes for offering these facilities.
The computation of the eigenvalues and the fitting was done
using routines from the GNU scientific library \cite{gnulib}.
I like to thank P. Majumdar for collaboration on earlier work
and discussing and reading early versions of this paper.
I also like to thank H. Wittig, H.B. Meyer and G. von Hippel
for fruitful discussions and reading the paper.
I receive support by the DFG via SFB 443.

\end{document}